# ANTS2 package: simulation and experimental data processing for Anger camera type detectors

A. Morozov[a,b,*], V. Solovov[a], R. Martins[a,b], F. Neves[a,b], V. Domingos[a] and V. Chepel[a,b]

[a] LIP-Coimbra,
Department of Physics, University of Coimbra, Coimbra, Portugal

[b] Department of Physics,
University of Coimbra, Coimbra, Portugal

[*] E-mail: andrei@coimbra.lip.pt

## Abstract

ANTS2 is a simulation and data processing package developed for position sensitive detectors with Anger camera type readout. The simulation module of ANTS2 is based on ROOT package from CERN, which is used to store the detector geometry and to perform 3D navigation. The module is capable of simulating particle sources, performing particle tracking, generating photons of primary and secondary scintillation, tracing optical photons and generating photosensor signals. The reconstruction module features several position reconstruction methods based on the statistical reconstruction algorithms (including GPU-based implementations), artificial neural networks and k-NN searches. The module can process simulated as well as imported experimental data containing photosensor signals. A custom library for B-spline parameterization of spatial response of photosensors is implemented which can be used to calculate and parameterize the spatial response of a detector. The package includes a graphical user interface with an extensive set of configuration, visualization and analysis tools. ANTS2 is being developed with the focus on the iterative (adaptive) reconstruction of the detector response using flood field irradiation data. The package is implemented in C++ programming language and it is a multiplatform, open source project.

## Table of Contents

# 1 Introduction

Many modern position sensitive particle detectors are based on the readout technique originally introduced by H.O. Anger to determine gamma-ray interaction position in a thin scintillation crystal for medical imaging applications [1]. In this technique, the scintillation is detected with a two-dimensional array of photomultiplier tubes (PMT) placed on top of the crystal. The scintillation position is determined from the distribution of signal amplitudes between the PMTs using, originally, the centre-of-gravity algorithm (COG, or centroid). Owing to its simplicity, reliability and good performance, the method introduced by H.O. Anger has found application in various types of radiation detectors used in different fields such as nuclear medicine (gamma cameras, see, e.g., review [2]), gamma-ray astronomy (e.g. [3]), imaging with thermal neutrons (e.g. [4] [5]), and dark matter search (e.g. [6] [7]). The origin of the scintillation photons is not very important and can be either primary or secondary scintillation (or both). Generically, we shall designate as *Anger camera type detector* any detector using arrays of photosensors to determine position of a source of scintillation photons produced as a result of particle interaction with the detector which can be considered as point-like and isotropic.

In medical imaging with radioisotopes, Anger camera with PMTs as photosensors remains the most widely used tool for two-dimensional or 3D (SPECT) imaging [2]. In the applications requiring compact cameras, PMTs are being replaced with silicon photosensors, among which silicon photomultripliers (SiPM) operated in the Geiger discharge mode are the most successful. A combined probe for prostate imaging is an example [8]. Monolithic scintillator detectors with SiPM readout are also being developed for SPECT and PET (e.g. [9] [10] [11] [12]). In some situations, using a multianode position sensitive PMT can be a good approach (e.g. [2] [3]).

The Anger camera type readout scheme was also implemented in gaseous and double phase (liquid/gas) detectors with secondary scintillation. In these detectors, the scintillation photons are emitted by gas atoms which are excited by the ionization electrons drifting in a strong electric field. Examples of such detectors using PMTs as photosensors are the double phase detectors in dark matter search (e.g. [13] [14]), high pressure and double phase xenon gamma-ray detectors that have been developed for medical imaging (e.g. [15] [16] [17]), and the microstrip-based thermal neutron detector with optical read-out [5]. In the NEXT detector, developed for neutrinoless double beta decay search, about 7000 silicon photodiodes are used in the tracking plane [18].

The performance parameters of an Anger camera-type detector, such as position and energy resolution, linearity, uniformity, size of the useful field-of-view, depend very much on the details of the detector design and its optimization. Optimization is a demanding task as it requires taking into account a large number of interconnected parameters. Also, the centroid position reconstruction algorithm (as originally employed by Anger) often does not provide sufficient image quality due to inherent distortions. Alternatively, statistical reconstruction algorithms (see, e.g., [19]) can be applied but then accurate knowledge of the detector spatial response is required. In turn, the response can also be optimized to satisfy certain criteria, for example to find a compromise between field-of-view uniformity and resolution. These issues can be addressed using numerical simulations.

A widespread approach for detector simulation is to use the Geant4 simulation toolkit [20]. Geant4 is capable of simulating all relevant processes of particle interaction with the detector components. However, due to the generality of the approach, the simulations tend to be slow and require deep knowledge of Geant4 itself. In what propagation and detection of optical photons is concerned, a number of optical parameters of the detector materials as well as detection properties of the individual photosensors should be known. Since such data for a real detector are often not available (the corresponding measurements are far from being straightforward), the predictive power of simulations, performed even with very detailed models, are usually quite limited. Geant4 also does not offer any tools for position reconstruction from the simulated energy depositions. Moreover, optimization of the detector performance and fine tuning of its parameters are difficult to perform with a universal package. The optimization process requires many successive simulations, often in the interactive mode, with variable geometry and different materials not to speak of details of the signal formation modelling.

These arguments have led to development of the predecessor of ANTS2, the software package ANTS [21] [22], capable to perform fast simulations of Anger camera type scintillation detectors. ANTS has been successfully applied for optimization of the microstrip-based gaseous secondary scintillation detector with an array of PMTs for thermal neutron imaging [5] [21]. The package allows to define simple planar detector geometries (i.e. composed by a combination of parallel slabs) with up to 100 PMTs or SiPMs. Along with the centroid position reconstruction algorithm, statistical reconstruction algorithms have been implemented. Additionally, a recently developed approach for calculation of the detector spatial response, based on an iterative procedure and using only flood field irradiation data (see, e.g., [23] [21] [24]), was introduced. The tolerance of the method to the precision of the initial guess on the detector response is quite high [21], that strongly reduces the requirements on the accuracy of the simulations from which the initial guess can be obtained.

One of the main advantages of that package is the short time (on the order of minutes) required to perform a complete optimization cycle, starting with detector geometry adjustment, simulation of a large set of events, and position reconstruction. This time scale is far beyond the capabilities of any tool based on Geant4. Taking into account that a large number of such cycles can be required to find the optimal response reconstruction procedure and, in some cases, to fine-tune the detector configuration, the short time scale is a crucial factor. Despite the limitations of the detector model implemented in ANTS, the simulated detector response data provided sufficiently good seed for a successful iterative procedure. The obtained light response functions (LRF) were used in statistical position reconstruction properly reproducing the respective detector irradiation pattern [21].

Due to the interest in the method from the community working with Anger camera type detectors and considering the limited capabilities of ANTS to handle detector types other than the thermal neutron detectors, development of ANTS2 package has been started. While keeping the capability to perform detector simulations starting from light emission from point sources inside the detector, the ANTS2 package is now able to simulate the entire process of signal formation in the detector. This includes interaction of a particle with the detector, propagation of the scintillation light, its detection with photosensors and generation of the detector output carrying information on the event position, time, and deposited energy.

Since scintillation detectors for medical imaging with gamma-rays and imaging with thermal neutrons were considered as the main application areas of ANTS2, interaction processes for these two particles have been primarily included. As the gamma-ray energy range of interest in medical imaging is of the order of tens to hundreds of keV, only photoelectric absorption and Compton scattering are considered. Pair production is not included in this version of ANTS2 as well as radiation losses for electrons. Compton and photoelectrons are considered to deposit their energy locally at the point of gamma-ray interaction. For many applications this is a sufficiently good approximation. For example, in the case of medical imaging with 140 keV gamma-rays of $^{99m}$Tc, the range of the electrons in NaI(Tl) is ≈0.1 mm, at most, while the radiation losses are <1%. In PET, scintillators with higher Z and density are usually used. In BGO, for instance, the range of the electrons due to 511 keV gamma-rays is <0.4 mm and the radiation losses are ~3%. If higher gamma-ray energies (or very high spatial resolution) are of interest, the simplified approach implemented in the current version may be insufficient. In gaseous medium, the electron range cannot be neglected, too. It is planned to be taken into account in the next versions of ANTS2, however at the level of details as low as possible in order to keep the optimization cycle time short.

For neutrons, only capture is included which can be followed by emission of charged particles, gamma-rays or neutrons. Multiple scattering is not simulated. Charged particles are tracked using user-provided stopping power, which can be imported from the specialized databases.

The ANTS2 simulation module is based on the 3D geometry configuration and navigation tools provided in the ROOT package from CERN [25] [26], which allowed to remove many limitations on the detector geometry compared to ANTS. The graphical user interface of ANTS2 offers several approaches to configure detector geometry, making it a quick and straightforward procedure. A larger set of reconstruction algorithms is now provided, which can be applied to simulated as well as imported experimental data. Some of the statistical reconstruction methods are implemented on a graphics processing unit (GPU), which allowed to reach reconstruction rates on the order of one million events per second using a consumer-grade personal computer [24].

The focus and specific strength of the ANTS2 package is in the iterative reconstruction of the detector response: all modules are designed to provide effective tools for finding an optimized reconstruction procedure for a given detector. A custom library for detector response parameterization based on B-splines [27] is implemented. It is capable of handling light response functions of photosensors depending on one, two or three spatial coordinates, which is sufficient for virtually any Anger camera type detector. In this paper we present the general structure of the package and describe the most important modules and their operation. Discussion of the user interface is omitted since a detailed user guide can be accessed online [28].

## 2 Package overview

The ANTS2 package was created to provide a comprehensive toolkit for development and optimization of Anger camera type detectors, focusing, in particular, on position and spatial response reconstruction techniques, including the iterative response reconstruction mentioned above.

As shown in figure 1, the package consists of six main modules: the detector configurator, the simulation module, the loader of experimental data, the reconstruction module, the LRF module and the collection of analysis/visualization tools. Interaction between the modules is organized through two data containers. The first one (detector definitions) holds all configuration settings of the detector concerning the detector geometry, materials, optical sensors and readout electronics. The second one (event data hub) contains all data related to the scintillation events, such as photosensor signals and, when available, true positions and number of emitted photons for each event. The event data hub is also used to store the reconstructed parameters of the events.

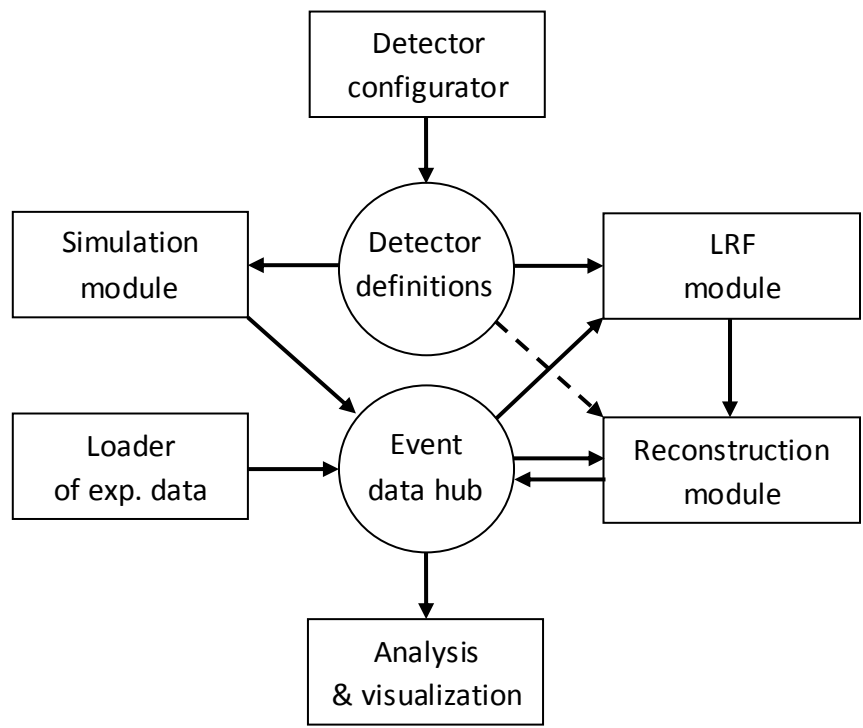

Figure 1. Diagram of ANTS2 general structure and data flow.

The detector configurator module allows the users to create or modify a model of an Anger camera type detector through a combination of a graphical interface and scripting tools. The user interface provides instant visual feedback for geometry creation/modification which makes this usually tedious process faster and less error-prone. The detector configuration can be saved, loaded or passed to the detector definitions data container using the JSON (JavaScript Object Notation) data-interchange format.

The simulation module performs Monte Carlo simulation of scintillation events in the detector based on the provided detector definitions. The module offers two operation modes: the *photon source* mode, in which simulation of every event starts with isotropic emission of optical photons at a given position, and the more advanced *particle source* mode, which involves simulation of particle emission from radiation sources and interaction of the particles with the detection medium followed by emission of scintillation photons. The output of the simulation module is a vector of the photosensor signals. These signals can be the number of detected photons or, if the parameters of the readout electronics are provided in the detector definitions, the amplitudes or digitized levels recorded by the data acquisition system. The signal data are stored in the event data hub alongside with the true values of the position and photon yield of each event. Additionally, a time-resolved simulation mode is provided which allows to simulate temporal evolution of the photosensor signals.

The loader of experimental data allows to fill the data hub with measured photosensor signals, and, for calibration datasets, the corresponding event positions/energies. It also offers some preprocessing tools: for example, a linear transformation (individual for each sensor) can be applied to the loaded signal values or some events can be filtered out according to user defined criteria.

The reconstruction module calculates spatial coordinates and energy (a quantity proportional to the number of emitted photons, to be more exact) of the scintillation events in the dataset with the photosensor signals stored in the event data hub. It is assumed that the light emission originates from isotropic, point-like sources. The module features several reconstruction algorithms: the centroid, statistical reconstruction, reconstruction based on artificial neural networks and nearest neighbour search (k-NN) based reconstruction. A set of event filters allows to discriminate events based on several parameters, for example, the reconstructed energy or, if available, the chi-squared of the reconstruction.

The statistical reconstruction requires knowledge of the light response functions (LRFs) that describe dependence of the photosensor signal on the position of a point-like isotropic light source in the detector. The LRF module calculates these functions from datasets (typically several tens of thousands events) containing photosensor signals and the corresponding event positions. The module can use the true positions of the events (known for simulation datasets or experimental calibration data) as well as the positions reconstructed by the reconstruction module. Cubic splines are employed for parameterization of the LRFs. A detailed description of the custom library used for parameterization and available parameterization options can be found in [27].

ANTS2 offers a set of tools for analysis and visualization of simulation results as well as the results of position, energy and detector spatial response reconstruction for simulated and loaded experimental data. A special focus is given to evaluation of the spatial/energy resolution and uniformity, which are the main criteria during optimization of the detector geometry and development of a response reconstruction procedure.

The package is implemented in C++ programming language as a multiplatform, open source project. The ROOT package from CERN [25], included as a library, is used to store 3D detector geometry and provides navigation during tracking of particles and photons. ROOT is also used in a number of computational tasks (e.g. minimization and fitting) as well as for visualization and data analysis. The graphical interface and the scripting engine were programmed using the Qt cross-platform application framework [29]. Several optional libraries can be included during ANTS2 compilation: the CUDA toolkit [30] from Nvidia to enable parallel calculations on GPU; the FANN library [31] for neural network-based position reconstruction and the FLANN library [32] for position reconstruction and event filtering based on k-NN algorithm.

## 3 Detector configuration

The ANTS2 package is developed aiming for both rapid prototyping during the design phase and in-detail studies of real detectors with complex configurations. In the first case, when many detector details can often be omitted, the highest priority is given to the simplicity of the detector model configuration and the instantaneous visual feedback. In the second case, the focus is shifted to the faithful reproduction of the properties of the detector. The configuration process becomes more complicated, which in part is mitigated by introduction of scripting facilities. With the ANTS2 package, a detector model can be developed stage by stage from a proof of concept to the final design by gradually adding the necessary details to the configuration.

### 3.1 Detector geometry (excluding photosensor arrays)

At the proof-of-concept stage of detector development, typically only the general shape has to be defined, which, in the case of an Anger camera type detector, often can be represented as a stack of slabs of different materials. Accordingly, the initial detector configuration can be simply defined by selecting the number of slabs in the stack and then choosing the material, thickness in Z-direction and the shape and dimensions in the XY plane for each slab. The shape (circular, rectangular or a polygon) can be defined as common for all the slabs in the stack or individually for each one. The stack is placed inside the "world" object, filled with a chosen material.

At the later stages, when finer details of the detector geometry have to be configured, the user can add additional objects (see [26] for the list of available shapes) to the geometry. These objects can be positioned at arbitrary coordinates and orientation as within the world object, as well as inside any object which has already been added to the geometry. Scripting tools are provided to facilitate configuration of objects with large number of daughter objects (e.g., complex lightguides or phantom objects, if an experiment is simulated).

Alternatively, as the most flexible approach, the detector geometry can be imported from a GDML file (XML-based geometry description [33]), which allows to introduce objects with a complex shape: see [26] for the list of available options. The detector geometry configured in ANTS2 can also be exported to a GDML file. Note that GDML geometry files can be generated (and read) by the Geant4 toolkit.

### 3.2 Photosensors

The photosensors are considered as interface between the optical and electrical parts of the detector and play a special role in ANTS2 simulation. As a part of detector geometry, they are characterized by their shape, dimensions and surface optical properties attributed to the photosensor material (e.g. the material of the photosensor entrance window). Additional properties are also defined to characterize the photoelectric conversion process, namely photon detection probability (optionally with wavelength, position and angular dependence), gain, single photoelectron response, and dark current.

Since in real detectors it is unusual to have more than a few different photosensor models, a concept of photosensor model was introduced in ANTS2 in order to allow sharing the same set of properties between several photosensors within one detector. While the geometrical parameters of each sensor are defined by the corresponding photosensor model, the photoelectric conversion properties can either be taken from the model definition or assigned to each photosensor individually. The first option is suitable for conceptual design stage while the second one is intended for simulation of real detectors when these properties can be obtained from calibration data.

ANTS2 can also simulate silicon photomultipliers: in this case the status of each individual Geiger-mode microcell is tracked during simulations in order to correctly reproduce the sensor saturation. Optional settings allow to configure electronic read-out properties (electronic noise and signal digitalization properties) for each individual photosensor.

### 3.3 Photosensor arrays

Up to two independent photosensor arrays can be configured in ANTS2. One of the three available options can be chosen for each array:

- A fully regular array in a plane perpendicular to the Z-axis with a square or hexagonal packing and a given inter-sensor distance;
- An array in a plane perpendicular to the Z-axis with custom XY positions of all sensors;
- An array with custom XYZ positions and orientations of all sensors.

For the first two options, the Z position of the sensor array is adjusted automatically to follow the defined detector slabs discussed in section 3.1. For the second and third options, the photosensor positions (XY and XYZ, respectively) are configured by the user with a script or by providing a file with the sensor coordinates.

### 3.4 Materials and particles

A material has to be associated with each object defined as a part of the detector geometry. It can be chosen from the internal material library or defined by the user. The material properties can be divided in two groups. The first group, described in this section, includes properties related to interaction of particles with matter. These properties have to be configured if simulations involving particle tracking are to be performed. The second group, described in the next section, contains optical properties, which are required in all simulation modes.

The configurable interaction properties depend on the type of the particle. In ANTS2 there are three types of particles which can be simulated: gamma-rays, charged particles and neutrons. Electrons are not included since for the targeted detector types it is typically possible to assume that electrons deposit their energy locally.

The interaction properties for each of these particle types are configured individually for each material. In the two extreme cases, the material can be set as either "transparent" or "opaque" for a given particle type. In transparent material these particles move without interaction, while upon entering an opaque material they are removed from tracking. By default, when no interaction properties are defined for a given particle type, the material is considered transparent for those particles.

Given that the gamma-ray energy range of interest is below 1 MeV, pair production is ignored and only photoelectric absorption and Compton scattering are considered. The user has to provide the mass attenuation coefficient as a function of energy for these two processes. Note that these data can be directly imported from files generated by XCOM [34].

For charged particles (these can be protons, deutrons, tritons, alpha particles, or heavier nuclei), the stopping power can be imported from a file generated by SRIM [35]. Straggling is not considered in ANTS2. Including this process would significantly slow down the simulations, while we expect negligible effect on the results relevant for the detector performance. For example, introduction of straggling had practically no impact on the position resolution obtained in simulations of thermal neutron detectors [22].

For thermal neutrons, the user has to provide the total capture cross-section and can configure the resulting nuclear reactions by providing their branching probabilities and defining the reaction products and their initial energies. Elastic scattering of neutrons in the materials normally used in thermal neutron detectors is, typically, weak and therefore is not considered in ANTS2.

For primary scintillation, the photon yield (i.e. the number of photons emitted per keV of deposited energy) should be defined for each relevant material-particle pair. If simulations involve secondary scintillation, the photon yield should be provided as the number of photons emitted per ionization electron; also, the average energy required to create an electron-ion pair has to be defined for the relevant materials.

For every material involved in simulations of particle interactions, the material density (as well as the isotope density, if neutron capture reaction is defined) has to be provided.

### 3.5 Optical properties and material interfaces

Optical properties of materials include refractive index, bulk absorption coefficient and Rayleigh scattering mean free path. For the materials, in which primary or secondary scintillation (or both) will be generated, the emission spectra and decay times can be defined. These properties are taken into account in the wavelength- and time-resolved simulation modes, respectively.

Generally, the behaviour of an optical photon at the boundary between two materials is completely defined in ANTS2 by the refractive indices of those materials and the photon incidence angle using Fresnel equations. However, for some interfaces, for example from gas to the detector wall or from the scintillator to the reflector, custom interaction rules can also be defined. ANTS2 provides a possibility to define probabilities of three additional processes at any given interface: absorption, specular reflection and diffuse scattering. For scattering, the Lambertian or isotropic profile can be selected.

# 4 Simulation module

The ANTS2 simulation module operates on the event-by-event basis. The event is defined as a complete sequence of the processes triggered by interaction of a particle with the detector, including emission of optical photons and the respective response of the photosensors. When the origin of optical photons is not important, the simulation of particle interaction can be omitted. In this case, we define event as a flash of light with a given number of photons emitted at a given point.

## 4.1 Simulation modes

The module offers two simulation modes: photon source mode and particle source mode. In the photon source mode, the user can define directly a source of optical photons for every event. In the particle mode, the user defines source(s) of particles. In this mode the number of generated photons is proportional to the energy deposited by the particle in the detector.

Simulations in the photon source mode are usually sufficient to generate data necessary for reconstruction of the spatial response of the detector with a high precision. Simulations in the particle source mode require significantly more configuration efforts, but are capable to provide more realistic output which can be directly compared with experimental data recorded with the corresponding detector.

### 4.1.1 Photon source mode

In this mode, simulation of an event starts with isotropic emission of optical photons at a given source position (node). The user chooses how many photons are generated: it can be a constant number or a random value with uniform, normal or user-provided distribution.

There are four event generation options for a simulation run (simulation of a given number of events):

- *Single node*: all events are generated with a fixed source position;
- *Regular scan*: source positions cover a user-defined regular grid;
- *Custom nodes*: the source positions are defined on event-by-event basis with a script or loaded from a file;
- *Flood field*: the source positions are randomly generated within the provided boundaries.

It is also possible to configure ANTS2 to perform a simulation run with a given number of events at each source position. In this case the user can access position resolution information provided by the dedicated tools of the reconstruction module.

As an advanced option, in order to characterize detector response to double interactions in the detector (Compton scattering, for example), the photons can be emitted simultaneously (within one event) from two nodes. The user defines the probability of such events and configures the number of photons emitted from each node.

It is also possible to select one of the two scintillation processes to be simulated: primary or secondary scintillation. For primary scintillation, all photons generated at a node are isotropically emitted from a point source. In the case of secondary scintillation, the photons are emitted from the points randomly distributed along the electron trajectory in the secondary scintillation region, assuming this trajectory to be oriented along the Z direction (see more details in section 4.2.2). To use this option, the secondary scintillator object should be defined in the detector geometry.

### 4.1.2 Particle source mode

In this mode simulation of each event starts from generation of a particle from one of the particle sources configured by the user. Each source has to be defined by providing its relative activity, particles which it can emit and their initial energies (fixed value or a custom energy spectrum). The user also configures the source shape (point, linear, area or 3D volume), size, position and orientation. In the case of non-point-like sources the activity is assumed to be distributed uniformly.

To simulate the collimation effect, emission of particles within a given cone, with the opening from 0 to 180 degrees and given orientation of its axis, can also be defined. The angular distribution within the cone is isotropic. Note that the sources are purely logical and there are no material objects associated with them in the detector geometry unless defined by the user. The proper configuration of the collimation options often allows to avoid introduction of special collimating objects in order to simulate particular experimental conditions.

It is also possible to allow generation of several particles per event: as mentioned above, this can be useful for optimization of event filtering for discrimination of multiple events.

## 4.2 Module operation

The flow diagram of the module operation in both simulation modes is shown in figure 2. In the particle source mode, for every event, the particle generator fills the particle stack using the user-defined configuration of the particle sources. The particles are then tracked and the information on the energy deposition is passed to the generator of primary or secondary scintillation. The generated photons are traced, the photosensor hits are detected, and the number of photoelectrons generated in each photosensor are recorded. Finally, the signal generator provides photosensor signal values, which are stored in the event data hub. In the photon source mode, the generated photons are passed to the photon tracer directly.

### 4.2.1 Particle tracking

The particles are tracked one at a time and the corresponding particle records are removed from the particle stack. If a new particle is generated during tracking (e.g. following a neutron capture), a new record is appended to the stack. The tracking is finished when the stack is empty.

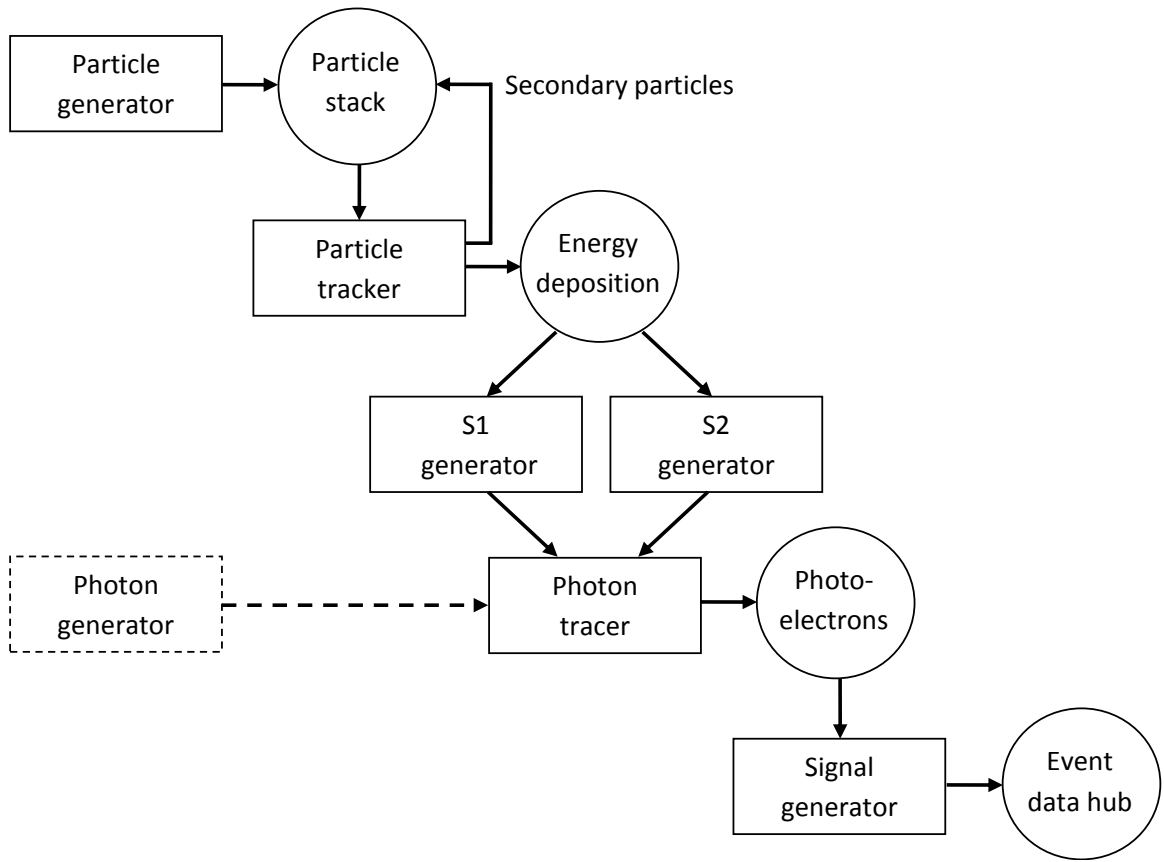

Figure 2. Flow chart of the simulation module operation in the particle sources mode. S1 and S2 are primary and secondary scintillation, respectively. The photon source simulation mode uses an alternative method of photon generation. Starting from the photon tracing phase, the operation is the same for both modes.

For charged particles, the stopping power in the current material is used. The stopping power table as function of particle energy can be imported, for example, from SRIM [35] or ASTAR/PSTAR [36] data bases or defined by the user. The step length is calculated according to the recommended fractional energy loss defined in the simulation settings. The step length is then clipped according to the user-defined minimum and maximum values. The energy loss is calculated and the corresponding record, containing the deposited energy, position, and the material/particle indices, is added to the energy deposition data (see figure 2). The procedure is repeated using new particle energy and position until one of the following situations occur: the energy reaches the cut-off value defined by the user, the particle exits the detector or it enters a material which is opaque for this particle type (see section 3.4).

For neutral particles (gamma-rays and neutrons), the mean free path in the current material is calculated for each interaction process. Then, random ranges corresponding to each process are generated and the one with the shortest range is selected. If this range is shorter than the distance to the next material border, the interaction is triggered. Otherwise, no interaction occurs and the particle is transferred to the next volume along its path. For gamma-rays, if photoelectric absorption is triggered, it is assumed that all energy is deposited at the interaction point. In the case of Compton scattering, the energy loss and new direction of the gamma-ray are randomly generated according to the Klein–Nishina equation. For neutrons, if capture is triggered, the records corresponding to the secondary particles are appended to the particle stack.

### 4.2.2 Photon generation

In order to generate scintillation photons, the energy deposition data are processed one record at a time. For primary scintillation, the photons are emitted isotropically from the position indicated in the record. The number of emitted photons is the product of the deposition energy and the photon yield for the corresponding material and particle. If multiple interactions occurred, the scintillation photons are generated at all interaction sites.

In the case of secondary scintillation, the procedure is more complex. Scintillation photons are produced in the volume, defined by the user as the secondary scintillator, if it is reached by the ionization electrons drifting in the electric field from the energy deposition site. The field in ANTS2 is always assumed to be antiparallel to the Z axis. The number of photons to generate is calculated according to the following equation:

$$N = \frac{\kappa E}{W}$$

where $\kappa$ is the number of photons per electron defined by the user for the material of the secondary scintillator, $E$ is the deposited energy and $W$ is the average energy required to create an electron-ion pair in the material indicated in the energy deposition record. The XY coordinates of the photon emission are given by the energy deposition position, while Z is randomly generated for each photon according to the uniform distribution within the Z-boundaries of the secondary scintillator at these XY coordinates. If multiple interactions occurred, this procedure is performed for all interaction sites.

For wavelength-resolved simulations, a random wavelength is assigned to each generated photon according to the scintillation spectrum defined for the corresponding material. In time-resolved simulations, each primary scintillation photon receives a random time stamp distributed exponentially with the decay time defined for the corresponding material. The time stamp for secondary scintillation photons is defined from the drift time of the generating electron from the point of origin to the emission point. The drift time is calculated as a sum of the drift times in each material that the electron encounters on its path. Electron drift velocities have to be defined for those materials. Electron attachment and diffusion are not simulated. The generated photons are passed, one at a time, to the photon tracer.

### 4.2.3 Photon tracing

Photon tracing is performed using the following elementary cycle. Starting at an interface between two materials (or at the photon emission position, if it is the first cycle), the position where the photon path intersects the next material interface is found according to the photon direction and the distance to that point is calculated.

Random ranges for absorption and Rayleigh scattering are found from the user-defined photon absorption coefficient and the scattering mean free path in the current material, respectively. If the shortest of the two ranges is within the distance to the next material interface, the corresponding process is triggered: for absorption, the photon tracing stops; for scattering a new photon direction is generated and the tracing continues.

If neither absorption nor Rayleigh scattering were triggered, the photon arrives at the material interface. In the case when the custom interface properties (absorption, specular or diffuse reflection - see section 3.5) are defined for the corresponding material pair, a random generator is used to determine which process is activated. If the custom properties are not defined, or none of the processes was triggered, the Fresnel equations are used to calculate the reflection coefficient assuming unpolarized light. According to this reflection coefficient, it is chosen whether the photon is reflected back or refracted to the next material. In the latter case the Snell's law is used to calculate the new photon direction.

The photon tracing stops if the photon is absorbed, leaves the detector geometry, enters a photosensor (the photon detection procedure is then triggered - see the next section) or a maximum number of photon tracing cycles, defined by the user, is reached. The limit for the number of cycles is implemented to avoid infinite tracing cycles, and a tool is provided to evaluate an adequate value using statistics on the number of tracing cycles for the detected photons.

All 3D navigation, including the determination of the object where the current position is located, finding the next material interface and calculation of the normal vector is performed using the TGeoManager class of the ROOT package [25] [26].

### 4.2.4  Photon detection

The detection probability of a photosensor is calculated as the product of three factors:

$$P_{det} = Q(\lambda) P_t(\Theta) P_A(x, y)$$

where $Q(\lambda)$ is the quantum efficiency of the photosensor at the photon wavelength, $P_t(\Theta)$ is the relative angular dependence of the detection probability as a function of angle of incidence (unity at the normal incidence), and $P_A(x, y)$ is the relative dependence of the detection probability (limited to the range from zero to unity) across the photosensor surface.

According to the detection probability, it is chosen whether the photon is detected or not using a random generator. For SiPM-type photosensors it is also checked which of the pixels (microcells) of the photosensor is triggered. Any repetitive hits of the same pixel in the same event are discarded. This check allows to properly simulate the limited dynamic range of this type of photosensors.

### 4.2.5  Photosensor signal generation

Depending on the purpose of simulation, the user can be interested in the raw number of detected photons for each photosensor, the signals generated by the photosensors or the values recorded by the digital acquisition system. By default, ANTS2 provides the number of the detected photons.

To simulate signal generation, in the simples case, the number of detected photons can be multiplied by a user-defined factor. This factor can be different for each photosensor, and thus be used to emulate differences in the gain of individual photosensors. It is also possible to simulate single photoelectron response of a photosensor: the contribution of every detected photon to the output signal can be drawn from normal, Gamma or a user-defined distribution.

Optionally, for SiPMs, the dark counts can be simulated using user-defined dark count rate and the acquisition time. It is also possible to simulate electronic noise by adding to each photosensor signal a normally distributed random value with zero average and the user-defined standard deviation. Signal digitalization noise of the readout electronics can also be simulated if the user provides the number of ADC bits and the dynamic range of the readout channels.

### 4.2.6 Acceleration options

In a typical Anger camera-type detector, the overall photon detection efficiency rarely exceeds 25%. Therefore, an optional photon tracing mode is provided which allows to considerably shorten the simulation time. Before tracing a photon, a random number is generated and compared with the computed beforehand threshold value equal to the maximum detection probability over all photosensors taking into account, if defined, the relative angular and area detection efficiencies. In the case when this random number is above the threshold, it is known that the photon, whatever photosensor it hits, will not be detected, and the photon is discarded without tracing. Otherwise, the standard photon tracing procedure is applied with the only modification being that if the photon hits a photosensor, the random number for the detection test (vs. the detection probability of this photosensor) was already generated before tracing. Note that this procedure is mathematically equivalent to the standard one, but allows to avoid unnecessary tracing of photons that are known not to produce a signal in photosensors.

The simulation module of ANTS2 is developed with multithreading capabilities. The user can configure how many CPU threads the simulation is allowed to use. Note that due to intensive usage of the computer memory during the simulations, the performance improvement is not scaling linearly with the number of threads.

## 4.3 Simulation output

After completion of a simulation, the photosensor signals as well as the positions and the number of emitted photons can be saved in a ROOT Tree file or exported to a text file. The sensor signal information can be accessed using ANTS2's GUI on event by event basis. It is also possible to perform 3D visualization of the photon and particle tracks superimposed on the 3D plot of the detector geometry.

Optionally, a detailed particle tracking and photon emission log can be generated. Users can also access the following information on the detected photons: photon wavelength spectrum, distribution of the angle of incidence at the photosensor, and distribution of the number of tracing cycles before detection.

# 5   Reconstruction module

## 5.1   Reconstruction methods

The reconstruction methods implemented in ANTS2 can be divided into four groups: centroid method, statistical algorithms, artificial neural networks and k-NN based reconstruction. The centroid reconstruction typically has low accuracy but it is a very robust technique which requires only information of the spatial positions of the photosensors, and, therefore, is very useful to make an approximate estimate of event positions. When the spatial response of the photosensors is known or experimental calibration data are available, statistical, ANN- and k-NN based reconstruction can also be performed. The statistical algorithms, being the most versatile, constitute the main workhorse of the package.

### 5.1.1 Centroid method

The centroid (or Center of Gravity, CoG) method is implemented in ANTS2 as follows:

$$x = \frac{\sum_i X_i S_i / g_i}{\sum_i S_i / g_i}, \qquad y = \frac{\sum_i Y_i S_i / g_i}{\sum_i S_i / g_i},$$

where *x* and *y* are the reconstructed positions, and ($X_i$, $Y_i$), $S_i$ and $g_i$ are the position, signal and relative gain of *i*-th photosensor, respectively. The relative gains can be directly provided by the user or estimated using a dataset of events recorded with flood field irradiation (see section 5.4). The centroid reconstruction can also be applied for Z-coordinate if the photosensors are not all positioned in a single plane.

Several options are offered to correct spatial distortions inherent to the method: the reconstructed positions appear "pulled" towards the detector center. A linear "stretch" can be applied independently in each direction: the reconstructed coordinates are multiplied by a given factor. Also, the photosensors situated far from the event position can be ignored during the reconstruction, which can cause the reconstructed position to shift closer to the true position. Selection of the sensors to be ignored can be based on the signal value and/or on the distance between the sensor position and that of the sensor with the highest signal.

### 5.1.2 Statistical methods

For each event, statistical methods search for the position (and, optionally, the event energy) which results in the best match between the simulated (or experimental) photosensor signals and those provided by a mathematical model of the detector. In ANTS2, the detector model is constructed using the light response functions of the photosensors.

All statistical methods implemented in ANTS2 can use either the Maximum Likelihood or Least Squares approach to evaluate the match between the sets of the photosensor signals. The best match position is found using either the algorithms provided in the Minuit2 minimization library [37] (included in the ROOT package) or the contracting grids [38] method. The Minuit2-based reconstruction is typically faster than the one based on the contracting grids (excluding the GPU-based implementation). However, the contracting grids method, when properly configured, has higher chances to converge to the global minimum.

It is also possible to configure reconstruction to ignore photosensors with the signals below a certain threshold or situated further than a given distance from the initial search position. This approach can improve accuracy of the position reconstruction by removing contribution from the photosensors with output signals dominated by noise.

#### *5.1.2.1 Minuit2-based reconstruction*

ANTS2 uses the Migrad and Simplex minimization algorithms implemented in Minuit2. These algorithms, according to our tests, proved to be the most robust, accurate and fast for the purpose of position reconstruction among all the other algorithms available in this library.

The user configures the selected algorithm to search for the minimum versus two (XY) or three (XYZ) spatial coordinates and, optionally, the event energy. Both algorithms are quite sensitive to the initial conditions, therefore ANTS2 offers two options for the initial search position: this can be the position reconstructed with the centroid method or just the coordinates of the photosensor with the strongest signal.

### *5.1.2.2 Contracting Grids*

ANTS2 uses a modified version of the contracting grids algorithm described in [38]. For each event, a regular grid of nodes is defined, centered at the start search position. The options for the start position are the same as for the Minuit2-based reconstruction. For each node, the algorithm evaluates the difference between the measured (or simulated) photosensor signals and the corresponding expected signal values given by the LRFs assuming that the light was emitted from this node. The position resulting in the best match is selected and used as the center of a finer grid with the same number of nodes. The procedure is repeated for the pre-configured number of iterations. The grid contraction factor and the number of iterations are typically chosen so that the step of the finest grid is significantly smaller than the spatial resolution of the detector. The event energy is evaluated at each node as the ratio of two sums over all photosensors: the sum of the photosensor signals and that of the LRF values [23].

One of the advantages of this algorithm is that it can be easily parallelized because evaluation for all the nodes of the same grid can be performed independently. Two implementations of the contracting grids algorithms are given in ANTS2: one version is executed on the CPU while the other one, exploiting this intrinsic parallelism, is coded using the CUDA platform [30] and is executed on the GPU of a CUDA-capable graphics card. Switching from the former to the latter can give a dramatic decrease in calculation time and, as demonstrated in [24], allows to reconstruct up to $10^6$ events per second for detectors with several tens of photosensors using a consumer grade Nvidia GeForce 770 graphics card.

### 5.1.3 Artificial neural networks (ANN)

The ANN-based reconstruction is performed in ANTS2 using the FANN [31] library. For a given detector, the user configures a neural network and then trains it using a calibration dataset with known event positions and energies. The calibration dataset should fully cover the range of event positions and energies for all experimental conditions for which the ANN-reconstruction is to be applied. After training, the network can be used to reconstruct event positions and energies.

ANTS2 implements almost all features available in the FANN library. Both fully connected and sparsely connected multilayer networks can be used. Several training methods are available, such as the standard back-propagation, adaptive back-propagation and the SARPROP algorithms [39].

For those users who have little experience with neural networks, the CASCADE2 method from FANN is implemented. This method automatically finds the best topology and properties of the network (e.g. the neuron activation functions and the number of neuron layers) and then trains the network using a calibration dataset.

During training of a network, the events are divided in two subsets: one is used for training, while the other one is reserved for monitoring of the performance of the trained network. The training proceeds automatically until the control parameters stabilize in either of the two subsets, which helps to avoid overtraining.

### 5.1.4 k-NN based reconstruction

A modified version of the k-NN based reconstruction algorithm described in [40] is implemented in ANTS2. The user has to provide a calibration dataset containing the photosensor signals and the corresponding event positions. The set should fully (and with a relatively high degree of uniformity) cover the field of view of the detector.

The algorithm operates in the N-dimensional space (N is equal to the number of photosensors), where each point represents a scintillation event and its coordinates are the normalized signals of the sensors. To reconstruct position of a scintillation event, the algorithm finds k calibration events (k is defined by the user) closest in this space to the point representing the reconstructed event. The XY position of the event is found as the centroid of the known positions of those k calibration events. The algorithm is implemented using the FLANN library [32] capable of performing efficient neighbour search in multidimensional space.

## 5.2 Event filtering

The reconstruction module features a set of event filters: individual events can be filtered out according to several criteria, including signal values of individual sensors or the sum signal of all sensors, reconstructed or loaded event energy, chi-squared value of the reconstruction, and the event position (true or reconstructed). These filters are typically used when background and double events have to be discriminated during the course of processing of experimental data or during analysis with the purpose of selection of a specific type of events. They can also be used in iterative reconstruction of detector response to discard events reconstructed with large errors (e.g. events with large chi-squared values or with unrealistic reconstructed energy or position).

A set of advanced filters allows to discriminate unwanted events using correlation of any two of the parameters listed above, or, based on a script, using custom event cuts joining multiple discrimination criteria. The advanced filters can be very useful, for example, for discrimination of signals due to interaction of gamma-rays in thermal neutron detectors.

Multiple filters can be activated at the same time. Only those events which pass all the activated filters are used in LRF reconstruction and are shown in any visualization of the reconstruction results. Well configured and optimized setting of the event filters is one of the key elements in finding an adequate iterative procedure of detector response reconstruction [24].

## 5.3 Passive photosensors

Passive status can be assigned to any number of photosensors: these photosensors are ignored during position reconstruction procedures. The basic usage of this option is obvious: one can disregard malfunctioning photosensors during reconstruction of experimental data or, with simulation datasets, to investigate the drop in performance of a particular reconstruction technique when one or more photosensors of the sensor array are not operational.

A possibility to disable individual photosensors during position reconstruction can also be very useful when response reconstruction procedure is performed (see section 8).

## 5.4 Gain evaluation

ANTS2 provides a set of tools to evaluate the relative gains of individual photosensors using flood field data (experimental or simulated, with events distributed approximately uniformly over the field of view of the detector) and taking advantage of the symmetry of the photosensor array. The most important methods, that have been already described in [22], are re-implemented in the ANTS2 package. The relative gains of all (or a set of) photosensors can be also evaluated using the LRF module (see section 6).

A preprocessing procedure can be applied to imported experimental data (see section 7) to take into account the evaluated gain values. After such correction, all operations within the reconstruction module can be done assuming equal gains for all photosensors.

## 5.5 Output

The reconstructed positions and energies can be saved in a ROOT Tree file or a text file. If available, the true event positions and the number of generated photons (or the loaded energy, if calibration data are processed) are included in both files. The Tree file also contains photomultiplier signals and the status of the event filters.

The reconstruction module offers an extensive set of tools for analysis and visualization of the reconstructed data. They can be easily accessed from the graphical interface, which is convenient for "on-the-fly" tweaking of the detector configuration or reconstruction settings as well as for analysis of the impact of the configuration changes on the detector performance.

For example, the user can interactively review the 3D positions of the reconstructed events superimposed on a wireframe 3D image of the detector geometry. It is also possible to produce 2D color-coded histograms of spatial distribution of the reconstructed event density, energy, chi-squared, sum signal of the photosensors and, for simulated or calibration data, deviations of the reconstructed positions from the true ones. For simulations performed in the photon sources mode with several events per node, the spatial resolution for each node can be reported and resolution versus XY position drawn as a color-coded map.

Using the Tree draw facility provided by the ROOT library, it is possible to draw distributions of many reconstruction and simulation-related parameters or to plot one parameter against any other (or two others) optionally applying an arbitrary number of parameter-based cuts.

## 6 LRF module

As it was discussed above, statistical reconstruction requires a mathematical model of the detector response. In ANTS2 this model is represented by a set of light response functions. An LRF describes the average signal of a photosensor as a function of the position of a point light source emitting isotropically a constant number of photons. The LRFs are parameterized in ANTS2 using B-splines [27]. The LRF module calculates these functions from datasets containing photosensor signals and the corresponding event positions.

## 6.1 LRF parameterization options

In general, an LRF is a function of three spatial coordinates (XYZ). However, in some cases one can successfully reduce its dimensionality to two or even one, that can significantly simplify the LRF parameterization procedures.

### 6.1.1 2D LRFs

For many position-sensitive detectors, only X and Y coordinates of an event have to be reconstructed (e.g. medical gamma cameras). In this case, it is often possible to ignore Z dependence of the LRFs. This type of parameterization scheme, when LRF is a function of X and Y and no other type of symmetry can be assumed, is designated in ANTS2 as "XY LRF". However, for some detector types the photosensor response has axial symmetry (see, e.g., [23] [21] [24]). In this case, LRF is a function of only one variable - the distance between the event position and the photosensor center in the XY plane. The corresponding parameterization scheme is designated in ANTS2 as "Axial LRF".

There is also the third case, when the photosensor response is predominantly axial, but due to light scattering there is a non-axial component in the response. For this case, ANTS2 offers the "Composite LRF" parameterization scheme. An LRF of this type is constructed as a sum of two components: an Axial LRF and an XY LRF. During LRF calculation, the axial component is evaluated first, and the residual is used to calculate the XY LRF component. This approach allows to minimize the required number of splines for adequate parameterization of the detector response which, in turn, results in a more compact set of parameters (important for GPU-based reconstruction using fast memory of a very limited size).

### 6.1.2 3D LRFs

Two parameterization schemes are available for detectors which require reconstruction of the source position in 3D. Using the first option, "Axial+Z LRF", it is assumed that the LRF has an axial response in XY plane and an independent variation along the Z direction.

The second option is "Sliced LRF", in which the LRF is represented by a set of several XY LRFs, each defined on a specific Z plane. The LRF values for the points between the planes are calculated using linear interpolation of the LRF values at two closest points from two closest planes.

## 6.2 Photosensor grouping options

Since Anger camera-type detectors are typically equipped with a large array of nearly identical photosensors, in some cases it can be advantageous to assign a number of photosensors to a group and use the same LRF for all photosensors within it. Since individual photosensors while having the same LRF profile can have different gains and quantum efficiencies, a scaling parameter (relative gain) has to be assigned to each of them.

When grouping is enabled in ANTS2, the following options are available:

- One group for all sensors;
- Grouping according to the distance from the sensor center to the detector center;
- Grouping based on rotation or/and reflection symmetry of the sensor array (square, rectangular or hexagonal)

Photosensor grouping allows to reduce the size of datasets required to accurately reconstruct the LRFs, since for reconstruction of a group LRF, every photosensor signal is used the number of times equal to the number of sensors in the group. The benefits of the sensor grouping for iterative LRF reconstruction procedure are discussed in section 8.

### 6.3 Operation and visualization

The LRF module allows to calculate light response functions using the selected parameterization scheme and grouping option. It also offers the possibility to re-calculate the relative gains of the photosensors within already defined groups without affecting the LRF profiles.

The user configures which type of the event position information is to be used for LRF calculation: the positions reconstructed by the reconstruction module, or the true event positions (in a simulation they are known; experimentally, they can be obtained from a calibration).

The calculated LRF for a given photosensor can be visualized from the GUI as a wireframe plot vs. XY coordinates or a radial plot (both with a possibility to overlay a scatter plot with signal values) so that the user can evaluate how well the LRF describes the photomultiplier signals within the provided dataset of the scintillation events.

## 7 Data import

Experimental data can be imported from a text file containing the photosensor signals (ordered by the photosensor number), one line per event. To load calibration datasets, the event position and energy (a value proportional to the total number of emitted photons) can be included at the end of each line.

When importing large sets of files (e.g., calibration runs with one file per source position), data can be loaded by providing a so-called manifest file. The manifest file contains the names of the files with the signal data and the position of the source (XY coordinates) for each file. As an option, it can also contain the shape and size of the collimated beam to be used in visualization of the reconstructed positions.

ANTS2 features several signal preprocessing options. For each individual photosensor, the user can configure a linear correction by providing the two coefficients. This feature can be used to subtract pedestal in experimental data and to account for relative gains of the photomultipliers. It is also possible to filter out events containing signals outside of a predefined range. This feature, for example, can be used to reject events with saturated signals.

## 8 Iterative LRF reconstruction in ANTS2

The method of iterative reconstruction of the spatial response of an Anger camera-type detector, previously described in [23], consists, in short, in the following procedure. Starting with an initial guess on the LRFs, the positions of the scintillation events recorded with flood irradiation are calculated with one of the statistical reconstruction techniques. In turn, the reconstructed positions and the measured photosensor signals are used to calculate the LRFs of the next iteration. The iterations are repeated until convergence of the LRFs is achieved. The initial LRFs can be directly calculated using simulation data (both the photosensor signals and the corresponding event positions are known in this case) or estimated from the measured photosensor signals and respective event position, reconstructed with the centroid method, which does not require knowledge of LRFs.

ANTS2 was developed with a special focus on this response reconstruction technique: the package provides a set of tools designed to assist users in configuration and optimization of the iterative procedure. The flexible approach to the LRF parameterization allows to minimize the number of parameters used to define an LRF profile, which can significantly increase the convergence rate and provides sufficient regularization of the LRFs. However, the user has to select the proper parameterization scheme and find the optimal set of the corresponding parameters to adequately parameterize the LRFs. Fast position reconstruction offered by the Minuit2-based, and especially the GPU-based implementations of the statistical reconstruction algorithms, is helpful here since it allows to quickly evaluate and compare the quality of the reconstruction performed with different LRF parameterization options.

While being quite tolerant to the initial guess [21], the iterative procedure nevertheless requires an adjustment for each particular detector. Several iterations at the very beginning of the procedure can be especially important: poor accuracy of position reconstruction at this phase can easily result in stagnation of the iterative process or cause a significant increase in the number of iterations required to reach the convergence. One of the solutions consists in grouping of peripheral photosensors with those situated close to the center of the sensor array since the latter ones usually have smaller errors in the reconstructed LRFs. Using a common LRF allows to establish an adequate general LRF profile for all the photosensors. After it is achieved, the grouping of the photosensors can be disabled and iterations continued to fine tune the individual LRFs. Also, application of a small random shift to the reconstructed positions usually improves the reconstructed LRFs during the initial iterations [24].

Sometimes the reconstructed image presents a stable artefact (i.e. artefact persisting in several iterations). It is frequently due to the photosensor in the neighbourhood of the artefact, whose LRF is defined with large errors relatively to that of the adjacent sensors. A possible solution is to disable this sensor for one iteration cycle. In this way, a more accurate position reconstruction in its vicinity can often be achieved, which, in turn, will results in a more accurate LRF obtained during the next LRF calculation phase.

Typically, a fraction of events is reconstructed with large errors due to deviation of the LRFs from the true ones. These events can lead to significant distortion of the reconstructed LRF profiles and result in a failure of the iterative procedure. Therefore, it is very important to adequately configure the event filters in order to discriminate such events during the iterative procedure. The discrimination criteria can be, for example, a large chi-squared value of the reconstruction or unrealistic reconstructed energy/position.

Some detector configurations can require a large number of iterations (e.g. ~30, as in [24]). Therefore, ANTS2 offers scripting tools to perform the iterative procedure in automatic cycles. The user configures the required number of cycles, and during each of them has access to all the tools described in this section. Additionally, to monitor the convergence, it is possible to perform visualization of the reconstructed event density/energy/chi-squared maps after each iteration.

## 9 Conclusions and future work

The ANTS2 package can be especially useful in two areas: development of new detector concepts based on Anger camera-type readout and establishment of an adequate iterative procedure for determination of the detector spatial response from flood field calibration data.

Two studies have already been performed which can be considered as the preliminary validation of the package. The iterative reconstruction technique was successfully developed using ANTS2 for a clinical gamma camera, retrofitted with an acquisition system capable to read signals of individual PMTs [24]. In another study [41], ANTS2 simulation results were compared, in terms of the spatial resolution and LRF profiles, with the experimental results recorded at a workbench developed to emulate Anger camera type detectors. Dedicated validation studies are still to be performed, including a cross-comparison with the results obtained with the Geant4 toolkit.

Development of the ANTS2 package is still ongoing. We continue to add new features related to the position reconstruction methods, event filters, models of processes implemented in the simulation module and LRF parameterization schemes. We also welcome ideas from the users and invite them to join the development team.

### Acknowledgments

This work was carried out with financial support from Fundação para Ciência e Tecnologia (FCT) through the project grants PTDC/BBB-BMD/2395/2012 (co-financed with FEDER) and IF/00378/2013/CP1172/CT0001, as well as from Quadro de Referência Estratégica Nacional (QREN) in the framework of the project Rad4Life. The authors would like to thank Dr. Martin Jurkovič for introducing them to CUDA and sharing his code.